\documentstyle[12pt]{article}
\def\ie{{\em i.e.}}
\def\eg{{\em e.g.}}

\def\beq{\begin{equation}}
\def\eeq{\end{equation}}

\catcode`\@=11 

\def\lsim{\mathrel{\mathpalette\@versim<}}
\def\gsim{\mathrel{\mathpalette\@versim>}}
\def\@versim#1#2{\vcenter{\offinterlineskip
    \ialign{$\m@th#1\hfil##\hfil$\crcr#2\crcr\sim\crcr } }}
\def\etal{{\em et. al.}}
\def\JL{J. L. Lopez}
\def\DVN{D. V. Nanopoulos}

\def\t1{{\tilde 1}}

\def\eV{\,{\rm eV}}

\def\Gyr{\,{\rm Gyr}}

\def\to{\rightarrow}

\def\NPB#1#2#3{Nucl. Phys. B {\bf#1} (19#2) #3}
\def\PLB#1#2#3{Phys. Lett. B {\bf#1} (19#2) #3}

\def\MODA#1#2#3{Mod. Phys. Lett. A {\bf#1} (19#2) #3}

\begin{document}
\begin{flushright}
\baselineskip=12pt
{CERN-TH/95--6}\\
{CTP-TAMU-69/94}\\
{ACT-25/94}\\
{hep-ph/9501293}
\end{flushright}

\begin{center}
\vglue 1cm
{\Huge\bf A new cosmological constant model\\}
\vglue 1cm
{JORGE L. LOPEZ$^{1,2}$ and D.V. NANOPOULOS$^{1,2,3}$\\}
\vglue 0.5cm
{\em $^1$Center for Theoretical Physics, Department of Physics, Texas A\&M
University\\}
{\em College Station, TX 77843--4242, USA\\}
{\em $^2$Astroparticle Physics Group, Houston Advanced Research Center
(HARC)\\}
{\em The Mitchell Campus, The Woodlands, TX 77381, USA\\}
{\em $^3$CERN Theory Division, 1211 Geneva 23, Switzerland\\}
\end{center}

\vglue 1cm
\begin{abstract}
We propose a new cosmological model with a time-dependent cosmological constant
($\Lambda\propto 1/t^2$), which starting at the Planck time as
$\Lambda_{Pl}\sim M^2_{Pl}$, evolves to the present-day allowed value of
$\Lambda_0\sim10^{-120}M^2_{Pl}$. This scenario is supported by non-critical
string theory considerations. We compute the age of the Universe and the
time-dependence of the scale factor in this model, and find general agreement
with recent determinations of the Hubble parameter for substantial values of
$\Omega_{\rm \Lambda}$. This effectively low-density open Universe model
differs from the traditional cosmological constant model, and has observable
implications for particle physics and cosmology.
\end{abstract}
\vspace{1cm}
\begin{flushleft}
\baselineskip=12pt
{CERN-TH/95--6}\\
{CTP-TAMU-69/94}\\
{ACT-25/94}\\
January 1995
\end{flushleft}

\vfill\eject
\setcounter{page}{1}
\pagestyle{plain}
\baselineskip=14pt

The problem of the cosmological constant ($\Lambda$) has been with us for over
75 years, ever since Einstein introduced it in order to avoid an expanding
Universe. Of course, Einstein was misinformed, although on the other hand,
since ``whatever is not forbidden is mandatory" one can say that he acted
in a physically reasonable way. A lot of effort has been spent tackling this
monumental problem, and yet a universally acceptable solution is still lacking.
The basic problem is that the vacuum energy of (spontaneously broken) gauge
quantum field theories, when coupled to gravity is metamorphosed into a
cosmological constant which is usually much larger than the {\em presently}
allowed one.

In this short note we propose yet another cosmological constant model. While
our model contains seeds of some ideas that have been developed recently in the
framework of non-critical string theory \cite{Time}, it is motivated by more
general, phenomenologically oriented principles, as well as by the recent
determination of the Hubble parameter ($H_0$) by the Hubble Space Telescope.

Let us start from the Einstein-Friedmann equation of standard Big-Bang
Cosmology
\begin{equation}
H^2={8\pi G\over3}\,\rho-{k\over R^2}+{\Lambda\over3}\ ,
\label{Friedman}
\end{equation}
where $H=\dot R/R$ is the Hubble parameter, $R$ is the scale factor, $G$ is
Newton's constant, $\rho$ is the particle energy density, and $k=+1,-1,0$ is
the curvature parameter for a closed/open/flat Universe. The present day
values of the various parameters ($H_0\sim 1/t_0\sim100\,{\rm
km\,s^{-1}\,Mpc^{-1}}$, with $t_0\sim10\Gyr$ the age;
$T_0\approx2.73\,{\rm K}$, $\Lambda_0\lsim10^{-120}M^2_{Pl}$) are extremely
small or large, and make the standard Big-Bang Cosmology rather unnatural.
Inflation has been introduced to solve some of these problems with great
success: an ``instantaneous" blow up of the scale factor ($R_{\rm
after}\sim10^{28}R_{\rm before}$) creates a smooth, effectively ``flat"
(\ie, $k/R^2\to0\Leftrightarrow k_{\rm eff}=0$) Universe of the right size and
entropy to describe our Universe. This natural obliteration of the curvature
term is rather effective, and it is interesting to see whether the same can be
done to the cosmological ``constant" term. Let us suppose that
\begin{equation}
\Lambda={\Lambda_{Pl}\over (R/\ell_{Pl})^\alpha}\ ,
\label{Lambda-alpha}
\end{equation}
where $\Lambda_{Pl}\sim M^2_{Pl}$ is the ``natural" size of the cosmological
constant, $(R/\ell_{Pl})$ is the scale factor in units of the Planck length,
and $\alpha$ is a constant to be determined by the present upper bound on
$\Lambda$: $\Lambda_0\lsim10^{-120}M^2_{Pl}$, and the present value of $R$:
$R_0/\ell_{Pl}\sim ct_0/\ell_{Pl}\sim10^{60}$. That is $\alpha=2$, and
\begin{equation}
\Lambda={\Lambda_{Pl}\over (R/\ell_{Pl})^2}\propto{1\over R^2}\ ,
\label{Lambda-2}
\end{equation}
has the same $R$ dependence as the curvature term, and would give us today
an acceptable and perhaps even detectable cosmological ``constant".
Let us emphasize the effect of inflation by rewriting Eq.~(\ref{Friedman})
as $H^2\sim M^2_{Pl}\,(T/M_{Pl})^4+M^2_{Pl}\,(\ell_{Pl}/R)^2$. Before inflation
$T\sim M_{Pl}, R\sim\ell_{Pl}$ and both terms are comparable; after inflation
and reheating $T\sim M_{Pl}$, $R\sim10^{28}\ell_{Pl}$ and the cosmological
constant term is very small compared to the radiation term. In the subsequent
adiabatic expansion $R\propto 1/T$ and we can write
$H^2\sim M^2_{Pl}\,(T/M_{Pl})^4+M^2_{Pl}\,10^{-56}\,(T/M_{Pl})^2$. This
relation shows that the cosmological constant term becomes comparable to the
radiation term at $T\sim10^{-28}M_{Pl}\sim1\eV$, \ie, the usual early Universe
cosmology is undisturbed.

It is also useful to give the time dependence of $\Lambda$, since this avoids
keeping track of the rapid change in $R$ during the inflationary phase. An
ansatz similar to that in Eq.~(\ref{Lambda-alpha}) gives
\begin{equation}
\Lambda={\Lambda_{Pl}\over(t/t_{Pl})^2}\propto {1\over t^2}\ ,
\label{Lambda-t}
\end{equation}
implying that $R\propto t$ somehow (this relation is only valid at late
times; in the early Universe $R\propto t^{1/2}$).

The above analysis is purely phenomenological, but suggests strongly that we
should seek a fundamental theory which can produce such $R$ or $t$ dependence
of the cosmological constant. In fact, an example of such a theory in the
context of non-critical string theory already exists in the literature. In
Ref.~\cite{ABEN} it was found that string theory admits only a few
cosmologically interesting (\ie, expanding Universe) solutions, all leading
asymptotically to a ``vacuum" characterized by
\begin{equation}
R\propto t\ ,\quad \Lambda\propto e^\Phi\delta c\propto 1/t^2\propto 1/R^2\ ,
\label{ABEN-stuff}
\end{equation}
where $\Phi$ is the dilaton field, and the central charge deficit $\delta
c\not=0$ reflects the departure from criticality and gives rise to the
cosmological constant $\Lambda$. Moreover, the curvature (Ricci) scalar ${\cal
R}\propto 1/t^2$, and in suitable units ${\cal R}=\Lambda$. That is, even in a
flat universe ($k=0$) we have an effective curvature which ``fakes" the
cosmology of an {\em open} Universe, as we discuss below. In other words, the
detailed string equations of motion for
this cosmological vacuum reduce, sufficiently below the Planck scale, to an
{\em effective} Einstein-Friedmann equation (\ref{Friedman}) with the
phenomenologically desired $R$-dependent cosmological ``constant".

It should be stressed that the above cosmological string vacuum should be
viewed as the correct asymptotic vacuum of string theory, just as the Minkowski
vacuum is to ordinary point-particle quantum field theory. As such, we expect
the above discussed ($R$-dependent) relations to hold in any realistic string
vacuum where the full gravitational and matter multiplets are accounted for.
The picture presented in Ref.~\cite{ABEN} has been completed by the addition of
string dynamics in Ref.~\cite{EMN}, where it was argued that not only the
cosmological ``constant", but all ``constants" in nature (\eg, $c,\hbar$)
become time dependent. Moreover, the desired initial conditions
$\Lambda_{Pl}=M^2_{Pl}$ and $k=0$ are derived in this dynamical scenario
\cite{EMN}. Also, in this dynamical scenario one obtains $\Lambda\propto 1/R^2$
without an explicit reference to a time-dependent dilaton field \cite{EMN}.
This is important since, as is, our toy model above entails a possibly
troublesome time-dependent gauge coupling ($g\propto e^{\Phi/2}$). Our present
phenomenological approach will be incorporated in the more rigorous string
dynamics picture of Ref.~\cite{EMN} elsewhere~\cite{ELMN}.

\begin{figure}[p]
\vspace{6.5in}
\includegraphics{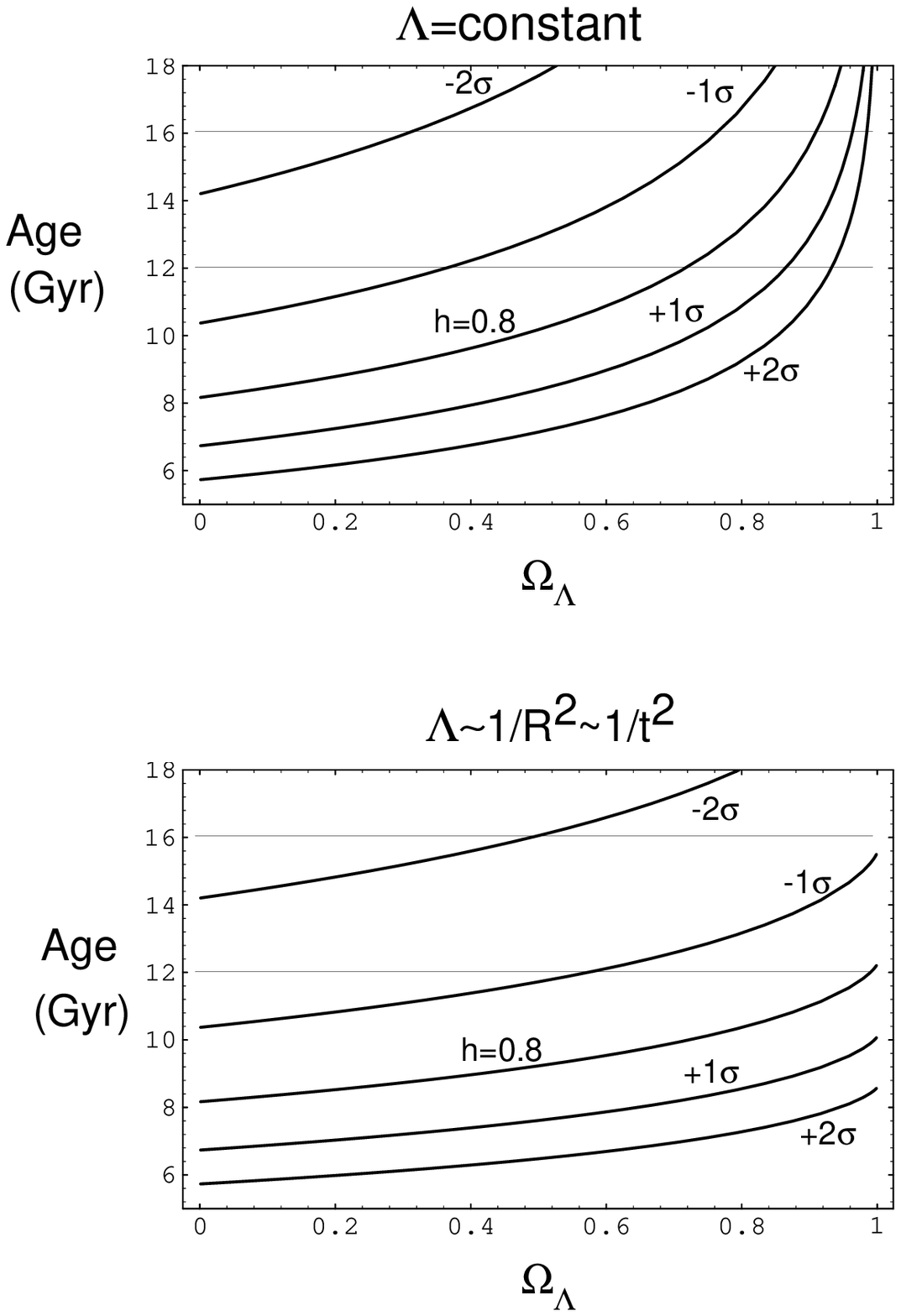}
\caption{The calculated age of the Universe (in Gyr) as a function of
$\Omega_{\rm \Lambda}$ in the traditional ($\Lambda=$constant) and our new
($\Lambda\propto1/R^2\propto1/t^2$) cosmological constant model, for various
choices of the Hubble parameter ($h=0.80\pm0.17$). The standard Big-Bang
Cosmology prediction (for $\Omega_0=1$) is recovered for $\Omega_{\rm
\Lambda}=0$. The horizontal lines delimit the estimated range for the age of
the Universe ($14\pm2\,{\rm Gyr}$).}
\label{Age}
\end{figure}

We now elaborate on the observational consequences of our cosmological model.
Starting from Eq.~(\ref{Friedman}), one can easily determine the age of the
Universe in the matter-dominated era, with $\rho=\rho^{\rm M}=\rho^{\rm M}_0
(R_0/R)^3$ and $\Lambda=\Lambda_0 (R_0/R)^2$ (also $k=0$). In our flat Universe
model $\Omega_{\rm M}+\Omega_{\rm \Lambda}=1$ (where $\Omega_{\rm M}=\rho^{\rm
M}_0/\rho_0$, $\Omega_{\rm \Lambda}=\Lambda_0/3H^2_0$, and
$H^2_0={8\pi\over3}G\rho_0$) and we get
\begin{equation}
t_0=H^{-1}_0 f(\Omega_{\rm \Lambda})\ ,\quad{\rm with}\quad
f(x)={1\over x}-{1-x\over x^{3/2}}\ln\left({1+\sqrt{x}\over\sqrt{1-x}}\right)\
{}.
\label{t0-us}
\end{equation}
This result has the property that $t_0$ is finite (\ie, $f(x)\le1$) for all
values of $\Omega_{\rm \Lambda}\le1$. For comparison, the traditional
cosmological constant model, where $\Lambda=\Lambda_0$ is truly constant, gives
\begin{equation}
t_0=H^{-1}_0 g(\Omega_{\rm \Lambda})\ ,\quad{\rm with}\quad
g(x)={2\over3}{1\over\sqrt{x}}\ln\left({1+\sqrt{x}\over\sqrt{1-x}}\right)\ ,
\label{t0-ccc}
\end{equation}
which diverges for $\Omega_{\rm \Lambda}\to1$. In numbers, the age of the
Universe is $t_0={1\over h}(9.8\Gyr) f(\Omega_{\rm \Lambda})$, where we take
the recent Hubble Space Telescope determination of $h=0.80\pm0.17$
\cite{Freedman}. The results for various values of $h$ and both cosmological
constant models are shown in Fig.~\ref{Age}. We note that for sufficiently high
values of $\Omega_{\rm \Lambda}$, both models are generally consistent with the
estimated age range of $14\pm2\Gyr$, although $\Omega_{\rm \Lambda}$ needs to
be larger in our new model. For instance, if $h=0.8$, $t_0>10\Gyr$ for
$\Omega_{\rm \Lambda}\gsim0.7$ in our model, whereas $\Omega_{\rm
\Lambda}\gsim0.45$ would do in the traditional model.

\begin{figure}[p]
\vspace{6.5in}
\includegraphics{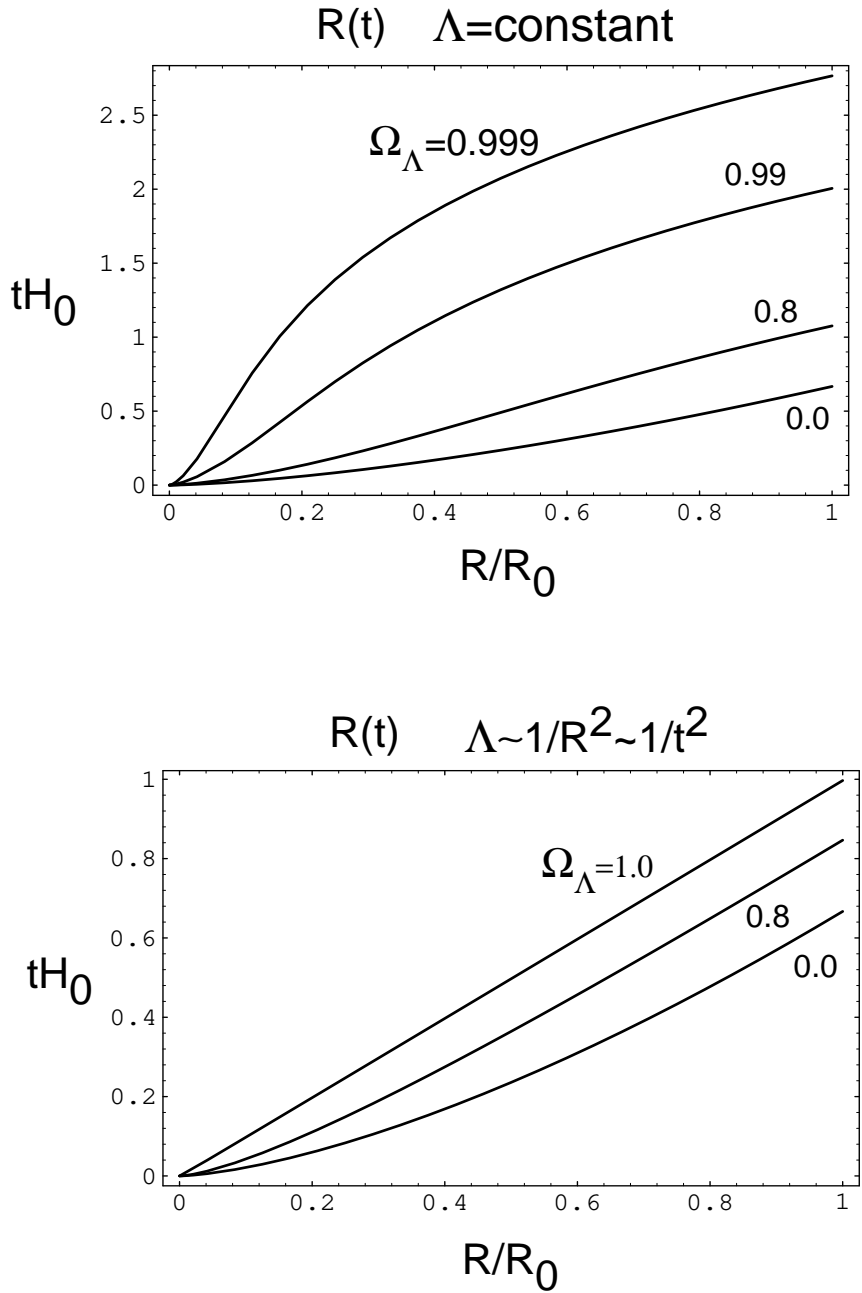}
\caption{The time dependence of the scale factor $R$ in the traditional
($\Lambda=$constant) and our new ($\Lambda\propto1/R^2\propto1/t^2$)
cosmological constant model, for various values of $\Omega_{\rm \Lambda}$. Note
that $R(t)\propto t$ in the new model for $\Omega_{\rm \Lambda}\sim1$.}
\label{R(t)}
\end{figure}

By integrating the Einstein-Friedmann equation up to an arbitrary time $t$, we
can also determine the time-dependence of the scale factor, which for an
``empty" (although ``filled" with a $t$-dependent cosmological ``constant")
Universe is supposed to be $R\propto t$.  We obtain
\begin{equation}
\sqrt{\Omega_{\rm \Lambda}}\ tH_0=
\sqrt{x(a+x)}-a\ln\left({\sqrt{x}+\sqrt{a+x}\over\sqrt{a}}\right)\ ,
\label{R-us}
\end{equation}
where $x=R/R_0$ and $a=(1-\Omega_{\rm \Lambda})/\Omega_{\rm \Lambda}$. For
$\Omega_{\rm \Lambda}=1$ this reduces to $tH_0=R/R_0$ as anticipated. In
contrast, the result in
the traditional cosmological constant model is
\begin{equation}
\sqrt{\Omega_{\rm \Lambda}}\ tH_0=
{2\over3}\ln\left({\sqrt{x^3}+\sqrt{a+x^3}\over\sqrt{a}}\right)\ ,
\label{R-ccc}
\end{equation}
which for $\Omega_{\rm \Lambda}\to1$ gives $R\sim e^{H_0t}$ as expected. The
$t$-dependence of the scale factor for various values of $\Omega_{\rm \Lambda}$
is shown in Fig.~\ref{R(t)}.

Since the age of the Universe constraint appears to require a substantial
value of $\Omega_{\rm \Lambda}$ (\eg, $\Omega_{\rm \Lambda}\gsim0.7$), in our
flat model the matter relic abundance will need to be suppressed (\eg,
$\Omega_{\rm M}\lsim0.3$). This situation would appear to be similar to the
traditional cosmological constant model studied in the literature \cite{ESM}
which has some advantages and some disadvantages. However, this is not so,
since in our case we effectively have a low-density open universe, with
different (and perhaps more desirable) properties regarding structure formation
\cite{Spergel}. Moreover, the particle physics contribution to the relic
density ($\Omega_{\rm M}$) can be easily accomodated in a supersymmetric model
with say $\Omega_{\rm M}h^2\lsim0.3(0.8)^2\sim0.2$, as explored previously in
Ref.~\cite{CC}.

In sum, we have proposed a new cosmological model with a ``running"
cosmological constant which, starting at the Planck time at its natural value,
evolves with time as $1/t^2$ and attains a present-day value in agreement with
observations. This model is supported by non-critical string theory
considerations, which also provide a justification for the initial conditions
used. Our effectively low-density open Universe model should be testable in
ongoing particle physics experiments as well as through cosmological
observations of the age and structure of the Universe.
Curiously enough, if indeed it holds true that the cosmological constant is
both non-zero and time-dependent, it would fulfill Dirac's prophesy that there
are no small constants, simply the Universe is too old \cite{Dirac}, and at the
same time it would redeem Einstein from what he thought was the biggest blunder
of his life, \ie, the introduction of the cosmological constant, leaving only
the very minor blunder of thinking that a cosmological constant different from
zero was indeed a blunder!

\section*{Acknowledgments}
This work has been supported in part by DOE grant DE-FG05-91-ER-40633.

\newpage

\clearpage

\clearpage


\newpage

\begin{thebibliography}{99}
\bibitem{Time}For a recent review see \DVN, ``As time goes by ...",
CERN-TH.7260/94 (hep-th/9406016).
\bibitem{ABEN}I. Antoniadis, C. Bachas, J. Ellis, and \DVN, \PLB{211}{88}{393},
 \NPB{328}{89}{117}, and \PLB{257}{91}{278}; for a review see \DVN\ in
Proceedings of the International School of Astroparticle Physics, HARC,
January 1991, edited by \DVN\ (World Scientific, Singapore 1991), p.~183.
\bibitem{EMN} J. Ellis, N. Mavromatos, and \DVN, in {\em Recent Advances in
the Superworld}, Proceedings of the HARC Workshop, edited by \JL\ and \DVN\
(World Scientific, Singapore 1994), p. 3 and references therein
(hep-th/9311148).
\bibitem{ELMN}J. Ellis, \JL, N. Mavromatos, and \DVN, in preparation.
\bibitem{Freedman}W.~Freedman, \etal, Nature {\bf371} (1994) 757.
\bibitem{ESM}G. Efstathiou, W. Sutherland, and S. Maddox, Nature {\bf 348}, 705
(1990); for a recent review see \eg, J. Silk, to appear in Proceedings of the
Beyond the Standard Model IV Conference, Lake Tahoe, December 1994.
\bibitem{Spergel}For a recent review see \eg, M. Kamionkowski, IASSNS-HEP-94/56
(astro-ph/9407062) and references therein.
\bibitem{CC}\JL\ and \DVN, \MODA{9}{94}{2755}.
\bibitem{Dirac}P.~A.~M.~Dirac, Nature, {\bf139}, 323 (1937);
Proc.~Roy.~Soc.~{\bf A165}, 199 (1938).
\end{thebibliography}
\end{document}